\documentstyle[11pt,mrs2001,epsfig]{article}
\begin{document}
\title{The Search for Matter with Gravitational Lensing}

\author{J. WAMBSGANSS}
\affil{Universit\"at Potsdam,
Institut f\"ur Physik,
Am Neuen Palais 10,
14469 Potsdam,
Germany
}

\begin{abstract}
Gravitational lensing is a powerful tool to detect compact matter
on very different mass scales. 
Of particular importance is the fact that lensing
is sensitive to both luminous and dark matter alike.
Depending on the mass scale, all lensing effects are 
used in the search for matter:
offset in position, image distortion, magnification, and 
multiple images.  
Gravitational lens detections cover three main mass ranges:
roughly stellar mass, galaxy mass and galaxy cluster mass
scales, i.e. well known classes of objects. 
Various searches based on different techniques explored 
the frequency of compact objects over more than 15 orders
of magnitude, so far mostly providing null results  in 
mass ranges different from the ones just mentioned.
In particular, no population of ``compact dark objects'' could
be detected so far. 
Combined, the lensing results offer some interesting limits on 
the cosmological frequency of compact objects in the mass
interval $10^{-3} \le M/M_\odot \le 10^{15}$, unfortunately
still with some gaps in between. In the near future, further studies 
along these lines promise to fill the gaps and to push 
the limits further down, or they might even detect new object classes.

\end{abstract}

\section{(Relevant) Basics of Lensing }

\begin{figure}[tb]
\plotone{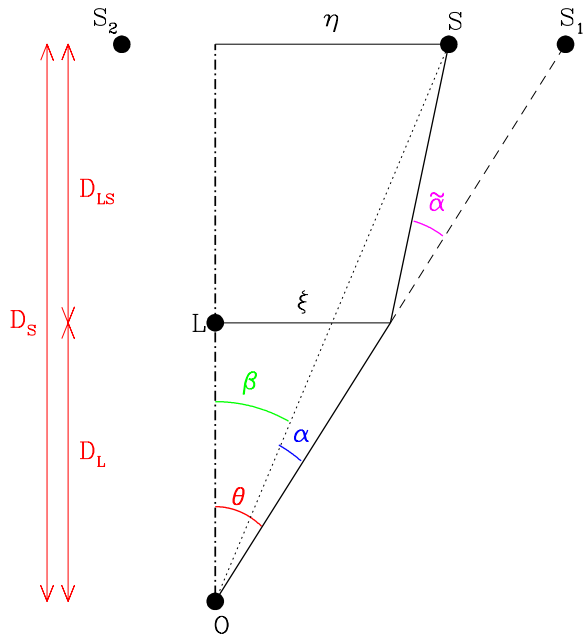}{100mm}
\caption{\label{fig-setup}
	a) Setup of a gravitational lens situation: The lens $L$ located
	between source $S$ and observer $O$ produces two images $S_1$
	and $S_2$ of the background source.
	Relations between the various angles and distances involved
	in the lensing setup can be derived for the case
	$\tilde \alpha \ll 1$, as 
	formulated in the lens equation (\ref{eq-lenseq}).
	}
\end{figure}

The basic setup for  a gravitational lens scenario is displayed in
Figure \ref{fig-setup}.
The three ingredients in such a lensing situation  are the
source S, the lens L, and the observer O.
Light rays emitted from the source are deflected by the lens.
For a point-like lens, there will always be (at least) two images
S$_1$ and S$_2$  of the source.  With external shear -- due to
the tidal field of objects outside but near the light bundles --
there can be more images.
The observer sees the images in directions corresponding to the
tangents of the incoming light paths.

In Figure \ref{fig-setup} the corresponding angles and angular
diameter distances $D_L$, $D_S$, $D_{LS}$ are
indicated.
In the thin-lens approximation,
the hyperbolic paths are approximated by their asymptotes.
In the circular-symmetric case, the deflection angle is
given as

\begin{equation}
\label{eq-angle}
\tilde \alpha (\xi)  =  { { 4 G M(\xi)} \over {c^2} } { 1  \over \xi }.
\end{equation}
where $M(\xi)$ is the mass inside a radius $\xi$. In this depiction
the origin is chosen at the observer.
   From the diagram it can be seen that the following relation holds:
\begin{equation}
                \theta D_S = \beta D_S + \tilde \alpha  D_{LS}
\end{equation}
(for $\theta$,  $\beta$, $\tilde \alpha \ll 1$;  this condition
is fulfilled
in practically all astrophysically relevant situations).
With the definition of the reduced deflection angle as
$\alpha (\theta)  =   ( D_{LS} / D_{S} ) \tilde \alpha (\theta)$,
this can be expressed as:
\begin{equation}
	\label{eq-lenseq}
       \beta =   \theta - \alpha (\theta).
\end{equation}
This relation between the positions of images and source
can easily be derived for a non-symmetric mass distribution as well.
In that case all angles are vector-valued.
The two-dimensional {\bf lens equation} then reads:
\begin{equation}
	\label{eq-lenseq-vec}
       \vec\beta =   \vec\theta - \vec\alpha (\vec\theta).
\end{equation}

For a point lens of mass $M$, the deflection angle is given
by equation (\ref{eq-angle}). Plugging this into equation (\ref{eq-lenseq})
and using the relation
$\xi =   D_L \theta $ (cf. Figure \ref{fig-setup}) one obtains:
\begin{equation}
       \beta(\theta)  =
       \theta - { D_{LS} \over D_L D_S} { 4 G M \over c^2 \theta}.
\end{equation}
For the special case in which the source lies exactly behind the
lens ($\beta = 0$), due to the symmetry a ring-like image occurs
whose angular radius is called {\bf Einstein radius} $\theta_E$:

\begin{equation}
       \theta_E  = \sqrt{ 			
			{ 4 G M \over c^2} 
			{D_{LS} \over D_L D_S}.
				}     
\end{equation}
The Einstein radius defines the angular scale for a lens
situation. For a massive galaxy with a mass of
$M = 10^{12} M_{\odot}$ at a redshift of $z_L = 0.5$
and a source at redshift $z_S = 2.0$
(we used here a Hubble constant of $H = 50 $km sec$^{-1}$ Mpc$^{-1}$ 
and an Einstein-deSitter universe),
the Einstein radius is
\begin{equation}
	\label{eq-angle-gal}
       \theta_{E, \ \rm galaxy}  \approx 1.8  \ \sqrt{ M \over 10^{12} M_\odot } \ 
	{\rm arcsec}
\end{equation}
(note that for cosmological distances in general
$D_{LS} \ne D_S - D_L$!).  For a galactic microlensing scenario in
which stars in the disk of the Milky Way act as lenses
for bulge stars close to the center of the Milky Way,
the scale defined by the Einstein radius is
\begin{equation}
       \theta_{E, \ \rm galactic\ star}  \approx 0.5  \ \sqrt{ M \over M_{\odot} }   \ {\rm 
milliarcsec}.
\end{equation}
Time scales for galactic microlensing events -- i.e. the duration for crossing the
Einstein radius -- typically range from weeks to months. 
For cosmological/quasar microlensing, this time scale extends to years; however,
caustic crossing events can be as short as a few weeks.

More detailed introductions to gravitational lensing including some historic
aspects can be found in \cite{livrev}, or 
in the textbook \cite{SEF} by Schneider et al. (1992) and in 
the more mathematically oriented monograph \cite{PLW} by 
Petters et al. (2001).

%
%
%
%

%
%
%
%
\section{Lensing Effects/Phenomena}

Light deflection/gravitational lensing has various effects on
background sources. 
Depending on the mass of the lensing object
(from comet-like objects to clusters of galaxies),
on the nature of the lensed source (point-like/unresolved or
extended), and on the detection (imaging and photometry in
the optical and radio regime, timing for gamma rays), the 
actual observations can cover quite a variety of techniques.
Here we consider only ``strong'' lensing, where the effect
can be seen for each case individually (for weak lensing,
see Yannick Mellier's contribution).
\subsection{How does matter affect the light of background sources?}
The consequences of strong lensing are:
\begin{itemize}
\item {\bf Change of position:} This is normally not observable, since 
	we do not have any information on the ``unlensed'' position of	
	a source; only in ``dynamical" situations,  in which
	the image/lens configuration changes with time, this can
	be observed (e.g., the sun passing in front of stars). Exactly
	this was, after all, the first detection of light deflection
	during the famous solar eclipse in 1919 \cite{dyson}.
\item {\bf Distortion:}
	The shape of resolved sources is changed by lensing. The best
	visualization of this effect are the giant luminous arcs.
\item {\bf (De)Magnification:}
	A few sources are (highly) magnified, most sources are
	slightly demagnified. This means that the
	luminosity function of a hypothetical population of
	cosmological ``standard candles'' will unavoidably 
	be broadened  (see, e.g.,  \cite{Wam97}).
\item {\bf Multiple images:}
	The most dramatic lensing phenomenon: multiple quasars and
	multiple galaxy images are observed directly, and
	via microlensing we have evidence of unresolved 
	multiple images as well. 

\end{itemize}

\begin{figure}[hbt]
\plottwo{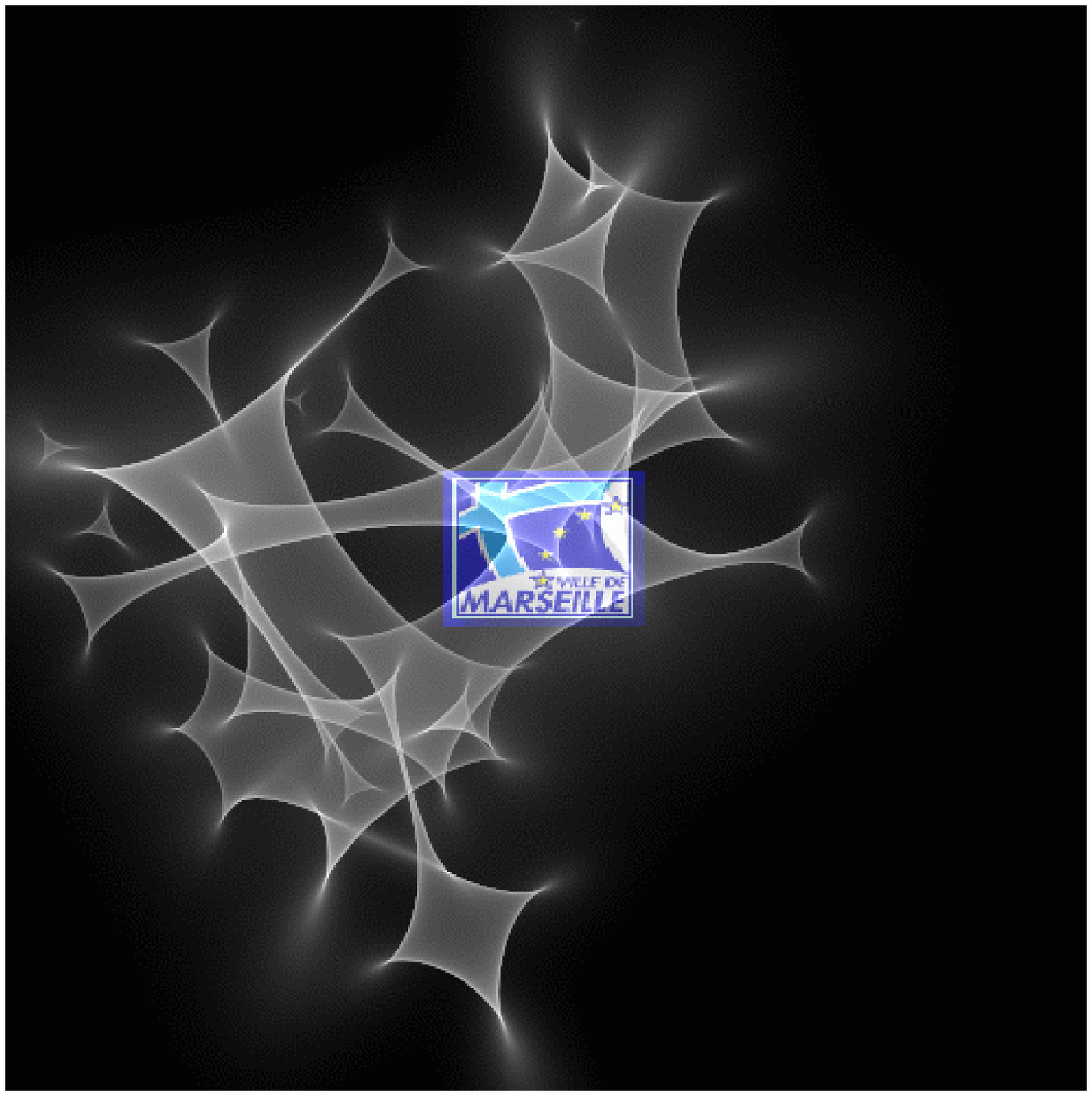}{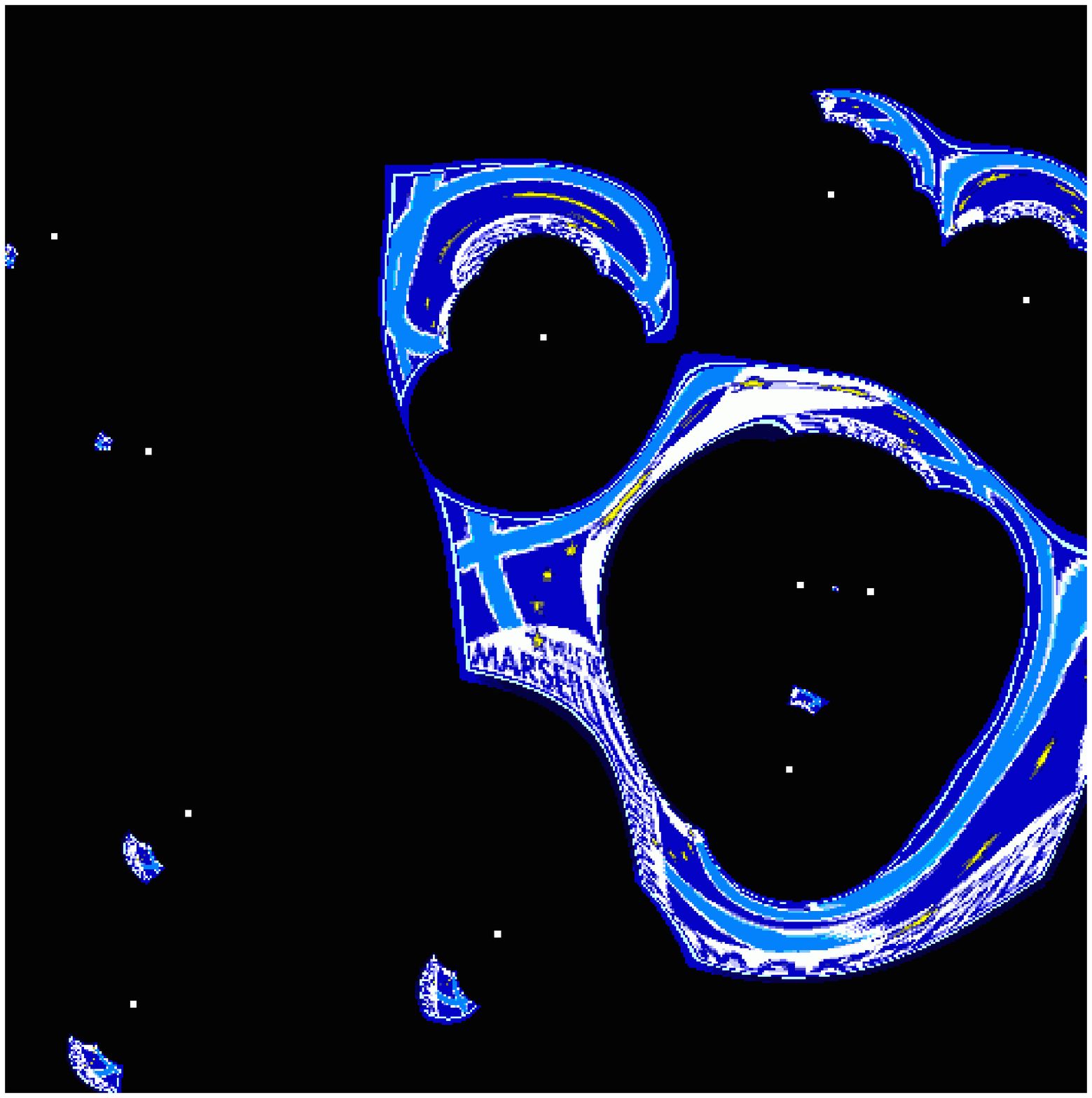}
\caption{\label{fig-ML}         
	a) Magnification distribution in source plane due to a number
	of point lenses (light grey means high magnification), 
	with specific source profile superimposed; 
	b) Corresponding image configuration  plus lens positions
	}
\end{figure}

\begin{figure}[hbt]
\plottwo {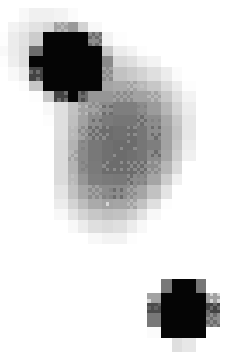} {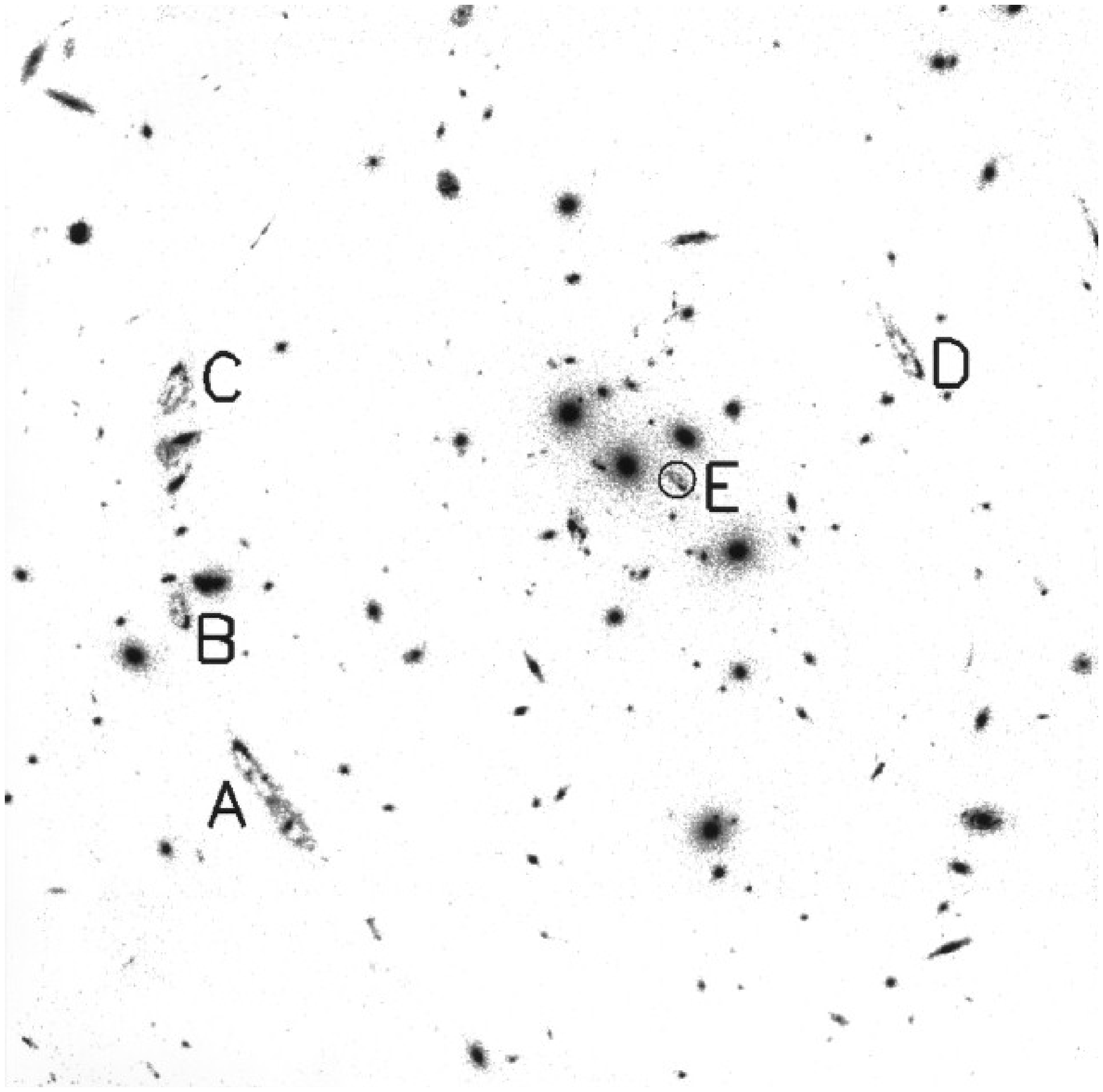}
\plottwo {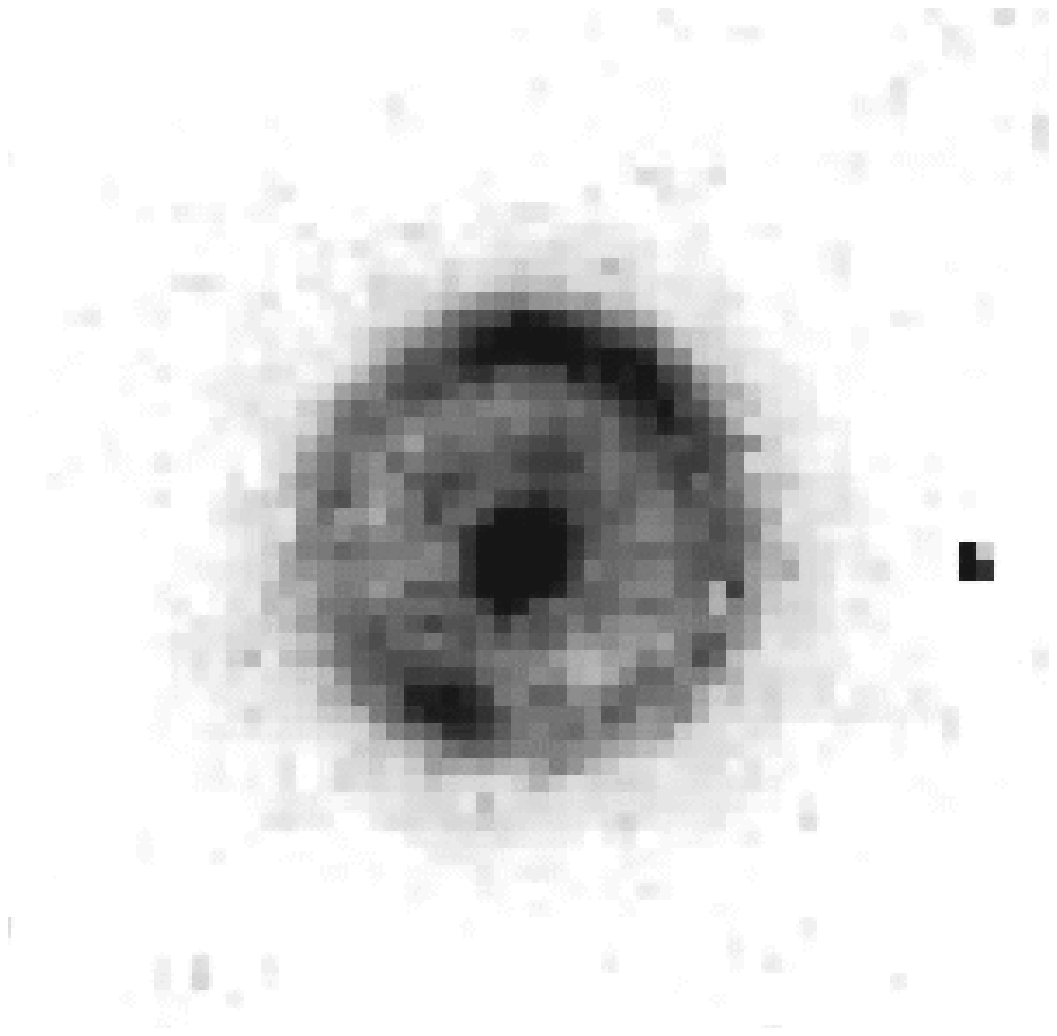} {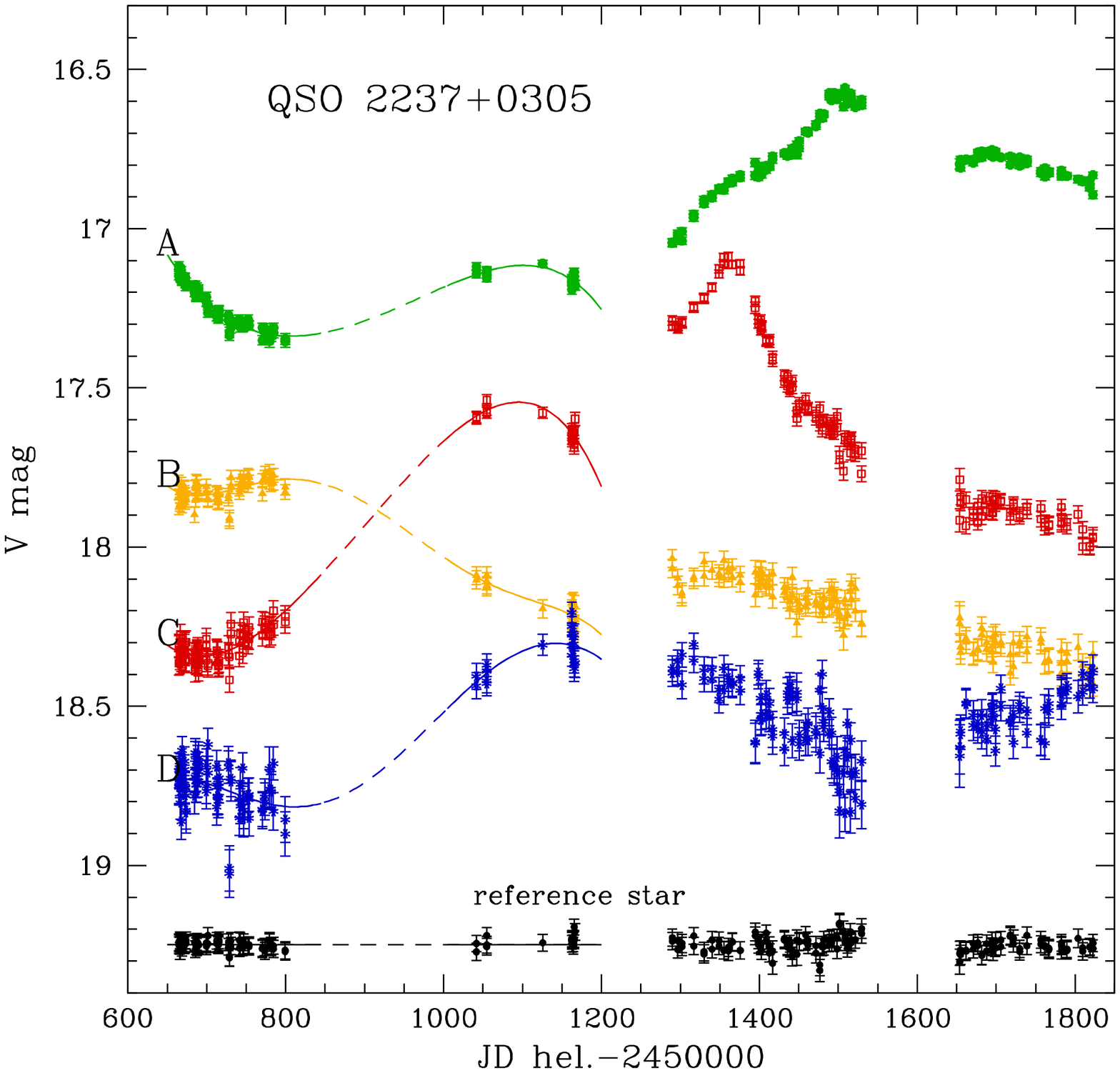}
\caption{\label{fig-4examples}         
	Four examples of strong lensing: 
	a) Double quasar HE1104-1805 (top left, \cite{courbin}):
        deconvolved infrared (J-band) image 
        of the two quasar images ($z_Q = 2.316$, $\Delta \theta =3.2$ arcsec)
	and the lensing galaxy  (at $z_G = 1.66$);
	b) Giant luminous arcs in cluster CL0024 (top right, \cite{CTT96}):
	five spectacular images of a high redshift galaxy seen lensed
	by a galaxy cluster (redshift $z_L = 0.39$) 
	with radius of curvature of about 20 arcseconds.
	c) Einstein ring B1938  (bottom left, see \cite{Kin97}): circular image
	with diameter 0.95 arcseconds;
	d) Microlensing in Q2237+0305 (bottom right, see \cite{wozniak1,wozniak2}):
	the lightcurves of the four images vary independently of each other,
	intrinsic variability can be excluded.
	}
\end{figure}

These effects often occur in combination. A slightly exaggerated
visualisation 
is provided in Figure \ref{fig-ML}: displaced, distorted,
(de-)magnified and multiple images of a source shape  with a particular
brightness profile. 
Quite a variety of spectacular lensing phenomena have been observed
in recent years.  
In Figure  \ref{fig-4examples},         
four of the most spectacular examples are presented: 
Multiply-imaged quasars, Giant luminous arcs,  Einstein rings
and quasar microlensing.

%
%
%
%

\subsection{Lensing phenomena:}
\label{sec-mult}

\subsubsection*{\underline{Multiple quasars}}
Multiply-imaged quasars were the first category of lensed systems
to be discovered \cite{walsh}.
By now more than 60 multiply-imaged quasar systems
have been found, most of them doubles or quadruplets,
recently even a six image configuration was discovered \cite{six}.
The angular separations range from a few tenth of an
arcsecond to about 8 arcseconds. The quasars are typically
at redshifts between $1.0 \le z_Q \le  4.5$. In almost
all cases, the lens is identified to be an intermediate
galaxy, in some cases ``assisted'' by a nearby group
of galaxies. 
Up-to-date tables of multiply-imaged
quasars and gravitational lens candidates
are provided, e.g., by the CASTLE group \cite{CASTLE}).

%
%
%
%

\subsubsection*{\underline{Einstein rings }}

A particular class of lenses are the Einstein rings, circular
images  of extended background sources. This scenario happens
when there is perfect alignment between source, lens and observer.
Since the radius of the Einstein ring is proportional to the
square root of the mass of the lens, these systems are
very good laboratories for weighing galaxies. 
The most remarkable example so far is the 
Einstein ring B1938+666 \cite{Kin97}. 
An infrared HST image shows an almost
perfectly circular ring with two bright parts plus the bright central
galaxy. 
By now about a half dozen cases have been found that qualify as Einstein 
rings \cite{CASTLE}.
Their diameters vary between 0.33 and about 2 arcseconds.
All of them are found in the radio regime, some have optical
or infrared counterparts as well.

\subsubsection*{\underline{Giant luminous arcs and arclets }}

Rich clusters of galaxies at redshifts beyond $z \approx 0.2$
with masses of  order $10^{14} M_{\odot}$ are very effective
lenses if they are centrally concentrated.
Since most clusters are not really spherical mass distributions
and since the alignment between lens and source is usually not perfect,
no complete Einstein rings have been found around clusters.
But there are many examples known of spectacular giant
luminous arcs which
are curved around the cluster center, with lengths up to about 20 
arcseconds. 
Their Einstein radii are of the order of 20 arcseconds, but
cases with radii up to  35 arcseconds are known \cite{schindler}.
One of the best known cases is the galaxy cluster 
CL0024+1654 (redshift $z = 0.39$), with  
red cluster galaxies and nicely elongated bluish 
arcs  \cite{CTT96}.
Images further out are less distorted, but still 
clearly visibly tangentially elongated: the arclets.
General results from the analysis of giant arcs and
arclets  in galaxy clusters are:
clusters of galaxies are dominated by dark matter, and 
typical ``mass-to-light ratios" 
for clusters obtained from strong (and weak, see below) lensing
analyses are \ \ $M/L \ge 100 \ M_\odot/L_\odot$.

\subsubsection*{\underline{Stellar and quasar microlensing} }
The lensing action of stellar mass objects is usually called
``microlensing''.
It comes in two varieties: star-star lensing, or ``local'' microlensing,
where stars in the Galactic disk or halo deflect the light
of background stars in the Galactic bulge  or in 
nearby galaxies  (LMC, SMC, M31).
The second variant is
star-quasar lensing, where stars in a distant (lensing) galaxy
act as microlenses on a quasar at cosmological distances.
In both cases, the action is measured as a characteristic light curve.

\subsubsection*{\underline{Further examples of lensing}}
``Millilensing'' has been proposed as well, for lensing objects
of roughly $10^6$M$_\odot$ objects, but has not been
observationally confirmed (see below).
Weak lensing, the tiny effects of galaxy clusters on 
background galaxies cannot been detected individually any more,
but due to the coherent tangential distortion a signal can
be measured when  the shapes  of 
thousands of background galaxies are analysed. ``Very weak lensing''
is used to measure the ``cosmic shear'', the effect of the
large scale structure of the universe on background galaxies.

%
%
%
%
\section{Tracing compact dark/bright matter with Lensing}
Gravitational lensing is a good means to
detect compact matter along the line of sight. ``Compact''
in this respect means: the size of the potential
lens has to be of order its Einstein radius  or smaller.
In practice, this means 
$$r < 0.02 (M/M_\odot)^{0.5} \rm pc,$$
which is easily fulfilled for stellar objects or galaxies.
Lensing can also detect 
gradients of the surface mass density,
i.e. smoothly varying surface mass density,
(cf.  Yannick Mellier's contribution on 
(very) weak lensing).
However, even a large amount of matter distributed
with a constant surface mass density over the sky would not
be detectable with point objects (it would affect the sizes 
of extended background objects, but in order to evaluate it,
this would require prior knowledge of unlensed sizes).

Lensing effects can be detected in two ways:
static measurements: (a) positions  (separations) and
shapes of objects are determined which do not change over
centuries or longer time scales;
and (b) dynamic measurements: changes in brightness or positions are
measured, usually on timescales of years or shorter.
In the following, we will discuss various mass regimes
with respect to what lensing can tell us about a
possible cosmic population. What is particularly important: 
No-shows matter!

\subsection{There are few (if any) Machos in galactic halos (or elsewhere)}

Paczy\'nski showed in 1986 that microlensing  can be used to
test whether the halo of the Milky Way is made of compact objects
of stellar or substellar mass \cite{Pac}. Occasionally, one
of the hypothetical objects should pass in front of
a background star in the Large Magellanic Cloud,
magnifying it in a very characteristic way. 
A few years later, three teams set out to measure this
effect: MACHO, EROS, and OGLE.
They did detect a number of events, but not as many as one would
expect, if the halo was made entirely of such Machos.
The latest results of MACHO and EROS are consistent with each other: 
after 5.7 years of MACHO monitoring of 12 million LMC stars,
13-17 events had been recorded (depending on the exact definition
of a microlensing  event). The conclusions are 
\cite{Alcock2000a}:
20\% of the Galactic halo could be made of objects in the
mass range $0.15 \le M/M_\odot \le 0.9$. The EROS team
arrived at similar conclusions: objects smaller than a few
solar masses are ruled out as important component of the
Galactic halo \cite{eros}. Due to the lack of 
long events (order years or longer), the MACHO team
could also put limits on black holes/dark matter objects
in the mass range $1.0 \le M/M_\odot \le 30$
\cite{Alcock2000c}.

Lensing can be used to test the compact population of
halos of other galaxies as well. In the double quasar
Q0957+561, image B is seen through the bright part
of the lensing galaxy, 1 arcsec off the center,
whereas image A is visible through the halo of this
galaxy, about 5 arcsec away from its center.
If the halo of this galaxy were made of Machos, then
the lightcurve of image A should be affected by occasional
microlensing and differ from the image B lightcurve.
Analysis of the two lightcurves (that
were originally measured in order to determine the time delay in
this system, see \cite{Kun97}) shows that they are
very similar. They are in fact not more different than
at most 0.05 mag, which is also the order of the observtional
uncertainty.
Comparison with numerical simulations shows that any population
of halo objects  in subsolar down to planetary mass range
should have produced larger differences (for moderately
small quasar sizes).  The conclusion is (see \cite{SW,w2000}):
it can be excluded that the halo of
this galaxy is made entirely of objects in the mass range
$10^{-6} \le M/M_\odot \le 10^{-1}$.

Summarized:
 LMC microlensing results AND quasar microlensing results confirm that halos
of galaxies cannot be made predominantly of Machos.

However, there is no doubt that lensing objects of stellar mass
exist:
towards the bulge of the Galaxy, a total of more than 
500 microlensing events has been detected, 
presumably due to low mass main sequence
stars \cite{Alcock2000b, eros, ogle}. 
This is a much higher
number of events than what was predicted from our knowledge of
the Milky  Way structure, and it still challenges some theoretical
models. 
The microlensing effect of stellar objects on a quasar has been detected as
well, most impressively in the quadruple quasar Q2237+0305 \cite{wozniak1,wozniak2}.  
Again, this can be explained entirely by an ordinary old population of
stars in the central parts of the lensing galaxy.

So far we only considered compact stellar mass objects bound in
halos of galaxies. However, it could be that there is a cosmological 
distribution of such objects. Press \& Gunn  \cite{PG}
showed in the early 1970s that gravitational lensing is an effective
method to detect such a population of condensed objects.
Dalcanton et al. (1994) did a study \cite{dalcanton} to search for 
such a cosmological population of objects in the 
mass range of $0.001 \le M/M_\odot \le 120$. 
They investigated the equivalent width distribution of 200 quasars,
based on the assumption that microlensing of such objects would
magnify the continuum emission of the quasars, because it emerges
from a much smaller spatial region than the broad and narrow line
regions. If microlensed, hence, this would reduce
the equivalent width of such affected quasars. In particular, one
would expect this to occur for quasars with higher redshift, since the
optical depth of such a distribution of compact objects would
increase with increasing source redshift.  No such effect
was found \cite{dalcanton}. Their conclusion was that 
$\Omega_{10^{-3} - 10^{1.3}} < 0.1$.  
More recently, it was claimed that
the variability of (single) quasars could be caused by
just  such a population of compact objects \cite{hawkins1,hawkins2},
however, this is still under debate and
the observational evidence is not yet conclusive.

\subsection{Few (if any) million solar mass black holes:
 	kinky VLBI jets in Q0957+561, 
	gamma-ray bursts, double radio sources}

There are various arguments that halos of galaxies could also
be made by black holes with masses around one million solar
masses.  This is an interesting mass range, because the
Einstein radius of such objects at cosmological distances
is of order a milliarcsecond, hence accessible to observations
with VLBI.   
And there are also objects out there to measure the effect:
the radio jets of lengths 50 to 100 milliarcseconds are
perfect targets for such a test. If there is a significant
population of lenses in this 
mass range between such a radio jet and the observer, they 
would produce bends and kinks and holes in such a jets.
The problem is that some/most jets have naturally 
bends and kinks, hence the lensing signature is not unique. 
However, nature provides us with a good test lab anyway: 
for multiply-imaged quasars, we have two or more images
of such a radio jet. And since this lensing effect
acts differently on either of these jets, we are able
to see whether such millilensing objects exist from comparing
the two radio jets \cite{WP92}.  In the case of the
double quasar Q0957+561, this test was done \cite{Garrett}.
The close similarity of the two jets excludes scenarios
in which more than 10\% of the halo is made of 
objects with $M > 3\times 10^6 M_\odot$.

A similar mass range for compact lensing objects can
be explored by searching for millilensed gamma-ray bursts.
Nemiroff et al. (2001) recently investigated 
774 BATSE-triggered GRBs for evidence of millilensing, i.e.
repeated peaks with similar light-curves
and spectra \cite{nemiroff}. 
Their null detection allowed them  to put limits on the
universal matter density in compact objects 
in the mass range 
$10^5 \le M/M_\odot \le 10^{9}$, excluding a significant 
population in this interval:
$\Omega_{10^{5} - 10^{9}} < 0.1$.

A study by Wilkinson et al. (2001) investigated 300 compact
radio sources with VLBI for possible double sources \cite{wilkinson}. 
They did not 
find any multiple images with angular  separations between
$1.5 \le \Delta \theta / \rm milliarcsec  \le 100$,
corresponding to a mass range of 
$10^6 \le M/M_\odot \le 10^{8}$.
They used this null result to put limits on the matter content 
in this form of supermassive objects:
$\Omega_{10^{6} - 10^{8}} < 0.01$  (2$\sigma$).

\subsection{Few (if any) $10^9$ - $10^{11}$ M$_\odot$ objects: no radio doubles}

Compact objects  in the mass range $10^9 M_\odot  - 10^{11} M_\odot$
have Einstein radii of a few hundredth to a few tenth of an 
arcsecond, an angular range difficult to access in the optical  regime.
However, the interferometric techniques in the radio make it possible
to probe it. Augusto \& Wilkinson \cite{AW01}
investigated 1665 sources
with a mean redshift of $<z> \approx 1.3$ 
(out of the more than 10000 objects in the JVAS/CLASS catalogue).
They searched with MERLIN for double images with angular separations 
between 90 and 300 milliarcseconds, corresponding to a mass range
of $3 \times 10^9 \le M/M_\odot \le 8 \times 10^{10}$. They did not
find a single lens, and their conclusion is that the
total matter in the universe in this mass interval cannot exceed
$\Omega_{10^{9.5} - 10^{10.9}} < 0.1$ (2$\sigma$).

\subsection{Many bright $10^{11}$ - $10^{13}$ M$_\odot$ objects: 
	64 multiple quasars cannot err!}

So far, 64 multiple quasars are known (see CASTLE web page, \cite{CASTLE}).
The separation distribution ranges from 0.33 arcsec to 6.93 arcsec.
The lens redshifts 
(31 of them measured) cover 
$0.04 \le z_{\rm Lens} \le 1.01$, whereas
the source redshifts (44 determined) span the interval from 
$0.96 \le z_{\rm Lens} \le 4.5$.
The mass range of these angular separations corresponds to  about
$10^{11} - 10^{13}$ solar masses, typical galaxy scales. Are
the lenses galaxies? Jackson et al. (1998) investigated
this question: in 12 out of 12 lens systems
which had originally been discovered in the radio regime, 
they found a galaxy in the optical or near-infrared \cite{Jackson}. So
the question ``are there any dark galaxy-mass lenses?'' was
solved three years ago: the answer was ``NO!''. However,
in the meantime the CLASS lens B0827+525 refuses to reveal
a visible lens galaxy, at least up to now. Koopmans (2000) 
called this system the best candidate for a ``dark lens'' \cite{koopmans}. 
So
the answer now is: ``MAYBE?''. But is is clear that if such
a population exists at all, it can make up only a
(very) small fraction of all the objects in this mass range.

\subsection{Giant arcs: lots of dark matter in galaxy clusters}

There are roughly 100 giant luminous arc systems known, highly
distorted background galaxies around clusters of galaxies. 
The most well known being
cluster 0024+1654 \cite{CTT96} and Abell 2218 \cite{Kneib}.
The radii of curvature of the arcs range from 10 arcsec to 35 arcsec 
\cite{schindler}.  In addition to these most dramatic arcs with
occasional length-to-width ratios of 10 or larger, there are 
numerous arclets and weakly distorted galaxies in these clusters
\cite{mellier}.
Techniques for cluster mass reconstruction provide excellent tools
to study the (total) mass distribution in clusters and compare with
the light distribution (see also contributions by Clowe and Lombardi).
The results show: galaxy clusters are dominated
by dark matter, consistent with studies based on velocity galaxy
dispersions or X-ray analyses.

The frequency of giant arcs can be used in a statistical sense
to constrain the underlying cosmological model, because the
various versions (flat matter-dominated; open; flat with large
cosmological constant) predict largely different arc abundances
\cite{bartelmann} (see also contribution by Meneghetti).

Searches for arcs are usually ``biased'': one looks for them
around known (massive) clusters of galaxies. So it is no surprise
that all known arcs systems are related to visible/bright galaxy 
clusters. Are there ways to test whether 
``dark clusters'' exist with masses in the range of galaxy clusters?
There are indeed, since they should produce
large separation multiple quasars as well. In a recent study,
Phillips et al. (2000, 2001) performed a careful study to search for
radio multiples with separations between 6 and 15 arcsec
\cite{PBWJ} as well as between 15 and 60 arcsec
\cite{PBW}.
In the former study, there remain $\le 1$ candidates
from 15000 flat-spectrum sources. 
In the latter investigation,
they found no radio multiples, and could provide
upper limits on any population of (dark) objects with 
$10^{13} \le M/M_\odot \le 10^{14} M_\odot$.
(The once so-called ``dark lens'' MG2016+112 turned
out to be a lensing
cluster of galaxies at 
a redshift of one, \cite{Soucail}).

The bottom line with respect to image splittings of order 30 to 70 arcseconds
is: 
there are lenses galore, i.e. galaxy clusters, but there is no evidence
for dark matter concentrations on these mass scales. 
(Already a decade ago, Nemiroff \cite{nemiroff91} excluded an $\Omega$-value
of more than 25\% in compact masses  between 
$10^{10}$ M$_\odot$ and $10^{15}$ M$_\odot$ from lensing studies.)

\subsection{Cosmological constant cannot be large}

Various authors showed in the late 1980s/early 1990s
that the frequency of multiple
quasar systems depends on the cosmological model 
(for a review see \cite{CTP}): the larger the contribution of
the cosmological constant the more multiple-imaged system one
expects. There were a few studies recently, which explored this
quantitatively. Based on a well-defined sample of
optical and radio lenses, Falco et al. (1998) concluded that 
the cosmological constant has to be smaller 
than $\Lambda < 0.62 \ \ (2\sigma)$,
in order to be consistent with the known frequency of lensed systems
\cite{FKM}.
Depending on the view, 
this is just about (in)consistent with values for the cosmological constant
as determined from recent supernovae searches at high redshift.
More work along these lines is clearly required.

%
%
%
%
\section{Dark matter -- bright prospects from Lensing?}
\begin{figure}[ht]
\vspace{2mm}
\plotone{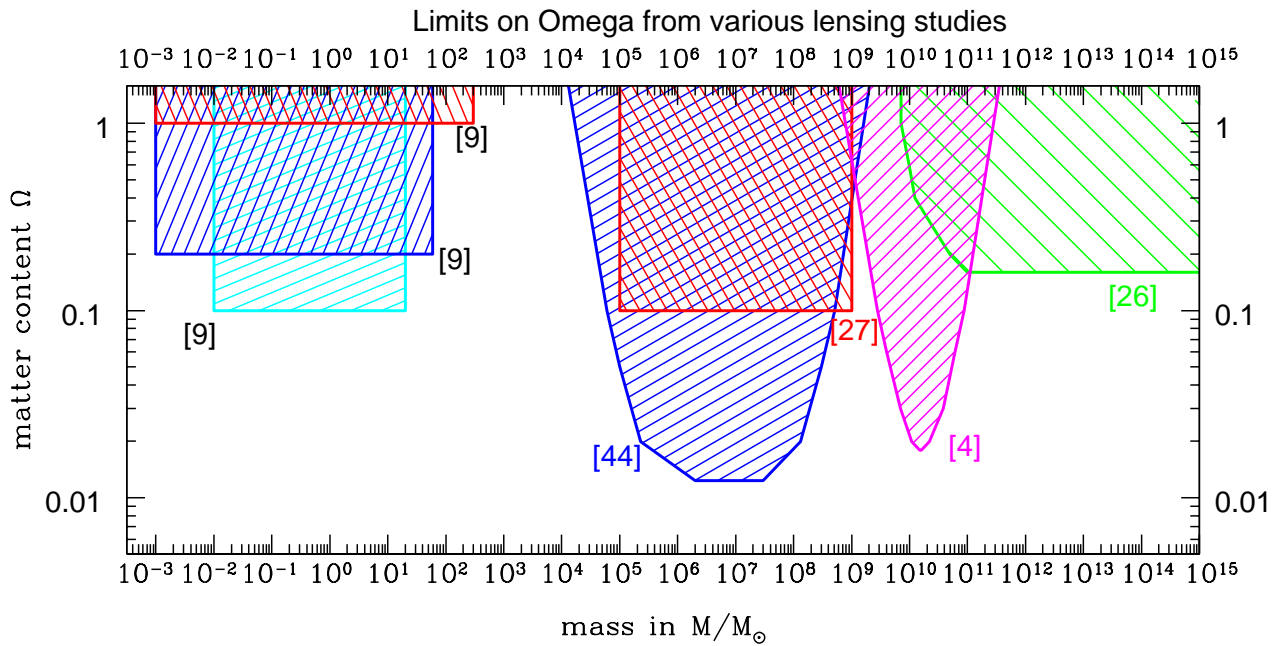}{170mm}
\vspace{-93mm}
\caption{\label{fig-omega}
	Limits on the matter
	content of the universe in form of cosmologically
	distributed compact objects:
	the shaded regions are excluded. 
	This diagram  combines
	various studies (as listed in brackets), 
	based on different techniques:  
	statistical microlensing of quasars
	\cite{dalcanton},
	VLBI investigation for multiple components of compact 
	radio sources \cite{AW01, wilkinson}, 
	frequency of  multiply imaged quasars \cite{nemiroff91}, or
	search for multiple gamma-ray bursts \cite{nemiroff}.
	}
\end{figure}

Strong lensing is a strong tool for the detection of compact 
matter in the universe. 
Studies searching for  lensing effects of compact objects
cover more than 15 orders of magnitude in mass:
from substellar objects
($\approx 10^{-2} M_\odot$) to galaxy clusters
($\approx 10^{14} M_\odot$), unfortunately still 
with a few gaps in between. 
A graphical summary of the results to date can be found in
Figure \ref{fig-omega}. 
%
%
%
%

Current gravitational lensing studies 
have good samples from lensing corresponding to mass scales
of (roughly) stellar mass, galaxy mass, and
galaxy cluster mass objects\footnote{In fact, even regimes
	covering mass ranges $10^{-16} \le M/M_\odot \le 10^{-6} $ are
	explored, so-called femto- and pico-lensing, see e.g. \cite{marani}}.
There is evidence for dark matter
in galaxies, and even more so in galaxy cluster. Few if any
really dark objects have been detected. Although there are
only upper limits on these dark lenses, it is obvious that they
cannot dominate the universe.

The big optical surveys underway (2dF, SDSS) will find many
more lens systems, based on well defined selection criteria 
which will provide much tighter limits on mass scales of
galaxies or larger. Radio surveys will provide data for
smaller separation/lower mass lenses. And new or more
``exotic'' aspects of lensing 
(astrometric microlensing measuring centroid shifts, gamma-ray burst
lensing measuring time delays) will bridge the gap between
mass scales of $\approx 10^{2} M_\odot$ and 
$\approx 10^{6} M_\odot$.
So as a matter of fact, prospects are very bright for   
more facts on (dark) matter.

\acknowledgements{It is a pleasure to thank the 
organizers Marie Treyer and Laurence Tresse and the LOC 
for organizing a great meeting and for generous support.}

\vfill
\end{document}